\documentclass[twocolumn,showpacs,preprintnumbers,amsmath,amssymb,superscriptaddress]{emulateapj}

\usepackage{graphicx}

\usepackage{aas_macros}
\usepackage{amsmath}
\usepackage{cases}

\usepackage{color}

\begin{document}


\title{Modeling the Anomalous
Microwave Emission with Spinning Nanoparticles: \\ No PAHs Required}%

\author{Brandon S. Hensley}
\email{brandon.s.hensley@jpl.nasa.gov}
\affiliation{Jet Propulsion Laboratory, California Institute of Technology, 4800
Oak Grove Drive, Pasadena, CA 91109, USA}

\author{B. T. Draine}
\affiliation{Department of Astrophysical Sciences,  Princeton
  University, Princeton, NJ 08544, USA}

\date{\today}

\begin{abstract}
In light of recent observational results indicating an apparent lack
of correlation between the Anomalous Microwave Emission (AME) and
mid-infrared emission from polycyclic aromatic hydrocarbons (PAHs), we assess
whether rotational emission from spinning silicate and/or iron
nanoparticles could account for the observed AME without violating
observational constraints on interstellar abundances,
ultraviolet extinction, and infrared emission. By modifying the \texttt{SpDust} code to compute the
rotational emission from these grains, we find that nanosilicate grains could account for the
entirety of the observed AME, whereas iron grains could be responsible
for only a fraction, even for extreme assumptions on the amount of
interstellar iron concentrated in ultrasmall iron
nanoparticles. Given the added complexity of contributions from
multiple grain populations to the total spinning dust emission, as
well as existing uncertainties due to the poorly-constrained grain
size, charge, and dipole moment distributions, we discuss generic, carrier-independent
predictions of spinning dust theory and observational tests
that could help identify the AME carrier(s). 
\end{abstract}

\section{Introduction}
Dust emission encodes information on the composition, size, and
temperature of the emitting grains and thus provides a window into the
evolution of heavy elements in the interstellar medium (ISM). A notable
historical example is the identification of the strong infrared
emission features at 3.3, 6.2, 7.7, 8.6, 11.3, 12.0, 12.7, and
13.55\,$\mu$m with emission from small
polycyclic aromatic hydrocarbons (PAHs) transiently heated to very
high temperatures \citep{Leger+Puget_1984,
  Allamandola+Tielens+Barker_1985}. PAHs, whose emission is ubiquitous throughout
the Milky Way and in external galaxies \citep{Smith+etal_2007}, are now thought to constitute roughly
5\% of the total Galactic dust mass and account for $\simeq 10$\% of the
interstellar carbon abundance \citep{Allamandola+Tielens+Barker_1989,
  Draine+Li_2007}.

The anomalous microwave emission (AME) is an emission component peaking near
30\,GHz, present in both the diffuse ISM and Galactic clouds, that is
strongly correlated with the far-infrared thermal
dust emission \citep{Dobler+Finkbeiner_2008a,
  Planck_Int_XV, Planck_2015_XXV}. The AME appears to be emission from
interstellar
grains and is therefore another observational window into their
properties.

\citet{Draine+Lazarian_1998a} proposed that the AME is electric dipole
radiation from rapidly-rotating ultrasmall grains, and this
explanation has gained wide acceptance due to its ability to account
for both the observed frequency-dependence of the emission
\citep[e.g.][]{Draine+Lazarian_1998a, Ysard+MivilleDeschenes+Verstraete_2010,
  Hoang+Lazarian+Draine_2011, Planck_Int_XV} and its
apparent lack of polarization \citep{GenovaSantos+etal_2015,
  Planck_2015_XXV}. In order to be driven to sufficiently
high rotation frequencies, the grains must be quite small (radius $a
\lesssim 10\,$\AA), leading to a natural association with the abundant
PAHs that give rise to the mid-infrared emission features. Indeed, it has been
demonstrated that both the mid-infrared emission features and AME can
be simultaneously accounted for by a population of PAHs with
reasonable assumptions on their abundance, size distribution, and electric dipole
moments \citep{Li+Draine_2001}.

Recent observations, however, have cast doubt on the association
between PAHs and the AME. In the Perseus molecular cloud
\citep{Tibbs+etal_2011}, the translucent cloud LDN\,1780
\citep{Vidal+etal_2011}, and the H{\sc ii} region RCW\,175
\citep{Tibbs+etal_2012, Battistelli+etal_2015}, the spatial morphology
of the AME does not match that of PAH emission at 8 and
12\,$\mu$m. Instead, the AME exhibits stronger correlation with
emission from very small grains at 25 and 
60\,$\mu$m, although in these dense regions interpretation is
complicated by attenuation of the starlight required to excite the PAH
emission. Resolved dust modeling of the
nearby spiral galaxy NGC\,6946 found a large scatter in the 30\,GHz
AME intensity per PAH surface density and no clear relationship
between the AME strength and the PAH abundance
\citep{Hensley+Murphy+Staguhn_2015}. Finally, a full-sky analysis
employing {\it Planck}
AME observations and 12\,$\mu$m WISE measurements of PAH emission found no
correlation between the strength of the AME and the abundance of
PAHs \citep{Hensley+Draine+Meisner_2016}.

Spinning dust theory predicts rotational emission from all
ultrasmall grains irrespective of composition as long as the spinning
grains have an electric or magnetic dipole moment. Given the apparent lack of
association between the AME and PAHs, it is natural to ask
whether the AME could be spinning dust emission from another
carrier. \citet{Hensley+Draine+Meisner_2016} suggested spinning
nanosilicates as a source for the
AME. \citet{Hoang+Vinh+QuynhLan_2016} found that spinning
nanosilicates can reproduce the AME if their
electric dipole moments are sufficiently
high ($\beta \gtrsim 0.2$\,D). \citet{Hoang+Lazarian_2016} found that metallic iron
nanoparticles could produce spinning dust emission with a 30\,GHz
emissivity within a factor of a few of the average Galactic value. However, they
calculated that these grains would be substantially aligned with the
interstellar magnetic field and thus violate the current upper limits
on AME polarization.

AME has been identified as a significant foreground for
upcoming, high-sensitivity CMB experiments aiming to detect primordial
B-mode polarization and spectral distortions. Correct modeling
of both the total intensity and polarization of this component is important
for unbiased recovery of cosmological parameters \citep[see,
e.g.,][]{Remazeilles+etal_2016}. While we have recently argued that
spinning dust emission should be effectively unpolarized due to
quantum mechanical suppression of the grain alignment process
\citep{Draine+Hensley_2016a}, it remains
unclear whether the AME is entirely spinning dust emission or whether
other emission mechanisms, such as thermal magnetic dipole emission, could be
acting in conjunction. A theoretical understanding of the spinning
dust SED, and its evolution with astrophysical environment, is thus
imperative.

The first aim of this work is to assess the viability of silicate and iron
nanoparticles as carriers of the AME. In particular, these grains,
like the PAHs, undergo stochastic heating and emit in the
mid-infrared. We thus ask whether a population of these grains can
simultaneously account for the AME without producing more infrared
emission than is observed. Likewise, we assess whether such grains
would produce more ultraviolet extinction than is observed. We find
that nanosilicates can reproduce the observed AME without violating
the observational constraints on mid-infrared emission and ultraviolet
extinction. The second aim of this work is to describe the observational predictions of
spinning dust theory that are carrier-independent. Within this more
generic framework, we suggest observational tests that would help
determine whether emission from spinning grains accounts for part or
the entirety of the observed AME.

This paper is organized as follows: in Section~\ref{sec:models}, we
specify the material properties of the silicate and iron nanoparticles
under assessment; in Section~\ref{sec:spdust},
we compute the spinning dust emission from populations of silicate and
iron nanoparticles for various assumptions on their size distribution;  in
Sections~\ref{sec:irem} and \ref{sec:uv_ext}, we compute the infrared
emission and ultraviolet extinction, respectively, from these
grains and compare to observational data; in
Section~\ref{sec:general}, we discuss carrier-independent
predictions of spinning dust emission and observational tests for
identifying the AME with a specific carrier;
in Section~\ref{sec:discussion}, we describe the
implications of our results on viability of non-PAH nanoparticles as
carriers of the AME; and we summarize our conclusions in
Section~\ref{sec:conclusions}.

\section{Potential Carriers}
\label{sec:models}
\subsection{Silicate Grains}
Silicon is highly depleted in the gas phase \citep[e.g.][]{Jenkins_2009}, and amorphous silicate
grains have been robustly identified as a major component of
interstellar dust on the basis of strong extinction features at 9.7
and 18\,$\mu$m. A significant population of sub-nanometer silicate
grains is a plausible component of the interstellar dust.

In this work, we assume that silicate grains have a chemical
composition of Mg$_{1.48}$Fe$_{0.32}$SiO$_{3.79}$ and a mass
density of 3.4\,g\,cm$^{-3}$ \citep{Poteet+Whittet+Draine_2015,
  Draine+Hensley_2016}.
We adopt an interstellar solid-phase silicon abundance of 40\,ppm in accord with
protostellar abundances and chemical enrichment
\citep{Chiappini+Romano+Matteucci_2003, Asplund+etal_2009} with 96\% of the Si
depleted onto dust \citep{Jenkins_2009} and define $Y_{\rm
  Si}$ as the fraction of the solid-phase interstellar silicon in nanosilicate
grains. We adopt the dielectric function for amorphous silicate of
\citet{Draine+Hensley_2016} and compute the grain temperature distributions
following the methods of \citet{Draine+Li_2001}.

As has been done with PAHs by previous studies, we model the electric
dipole moment of an amorphous silicate grain as the result of a random walk such
that the rms electric dipole moment scales as $\sqrt{N_{at}}$, where
$N_{at}$ is the number of atoms in the grain. Thus the probability of
a grain of $N_{at}$ atoms having an intrinsic electric dipole moment
$\mu_i$ is given by

\begin{equation}
\label{eq:dipole_dist}
{\rm d}P\left(\mu_i\right) \propto \mu_i^2 e^{-\mu_i^2/\left(\beta^2
    N_{at}\right)}\ {\rm d}\mu_i
~~~,
\end{equation}
where $\beta$ is the rms dipole moment per atom. 

In addition to the intrinsic dipole moment, we also consider a dipole
moment arising from the displacement of the grain's charge centroid
from its center of mass. Following \citet{Draine+Lazarian_1998b}, we
take the magnitude of this electric dipole moment to be $\epsilon a_x Z e$
where $Ze$ is the grain charge and $\epsilon a_x$ is the vector
displacement between the charge centroid and the center of mass. We
adopt $\epsilon = 0.01$ and, for the spherical grains considered in
this work, $a = a_x$ (note that disk-like grains such as small PAHs
have $a_x$ given by \citet{Draine+Lazarian_1998b}, Equation~3). Thus,
the total electric dipole moment is

\begin{equation}
\label{eq:e_dipole}
\mu^2 = \mu_i^2 + \left(\epsilon a_x Z e\right)^2
~~~.
\end{equation}

If the electric dipole moment and rotation axis are randomly oriented,
as might be expected in spherical grains, then the component of
$\mu$ along the rotation axis has an average magnitude of $\mu_\perp^2
= 2\mu^2/3$. To compute the spinning dust emissivity from a
population of grains with $N_{at}$ atoms, we average
Equation~\ref{eq:e_dipole} over the $\mu_i$ distribution of
Equation~\ref{eq:dipole_dist} following
\citet{Ali-Haimoud+Hirata+Dickinson_2009}. In this work we consider
$\beta = 0.3$ and $1.0\,{\rm D}$, with the former value
corresponding roughly to the value adopted for PAHs and the latter an estimate
based on various silicate materials
\citep[see][Table~1]{Hoang+Vinh+QuynhLan_2016}. 

\subsection{Iron Grains}
Like silicon, iron is heavily depleted in the gas phase
\citep[e.g.][]{Jenkins_2009} and is thus a major constituent of
interstellar dust. The presence of included iron nanoparticles in
lunar soil grains \citep{Keller+McKay_1997}, interplanetary dust
particles \citep{Bradley_1994}, and putative interstellar grains found in the
Solar System \citep{Westphal+etal_2014, Altobelli+etal_2016} suggest
that iron nanoparticles could be a significant constituent of
interstellar dust.

In this work, we consider populations of iron nanoparticles with mass density
$\rho = 7.87\,{\rm g}\,{\rm cm}^{-3}$. We adopt an interstellar iron abundance
of 41\,ppm in accord with observations of young F and G stars
\citep{Bensby+etal_2005, Lodders+etal_2009}, with 99\% of the
interstellar iron depleted onto grains \citep{Jenkins_2009}. We define
$Y_{\rm Fe}$ as the fraction of the solid-phase iron present in the
form of iron nanoparticles and adopt the metallic iron dielectric
function of \citet{Draine+Hensley_2013}.

\subsubsection{Magnetic Dipole Moment of Fe Nanoparticles}
Because metallic Fe is conductive, the electric dipole moment is
expected to be small for pure Fe nanoparticles. However, unlike
nanosilicates and PAHs, metallic iron nanoparticles are ferromagnetic
and can produce spinning dust emission due to their {\it magnetic}
dipole moment $\mu_m$ \citep{Hoang+Lazarian_2016}. The 
magnetization is ordered and therefore the total magnetic dipole
moment grows linearly with the number of iron atoms, in contrast to
the random walk process used to model the total electric dipole
moments of nanosilicates and PAHs. We therefore adopt

\begin{equation}
\label{eq:fe_mu}
\mu_m = 0.027 N_{at}\ {\rm D}
~~~,
\end{equation}
where we have assumed a magnetic dipole moment per atom of 3$\mu_{\rm B}$,
appropriate for small clusters of iron atoms
\citep[$N_{at} \lesssim$ 200,][]{Billas+etal_1993, Tiago+etal_2006}.

For these grains we also consider the small electric dipole moment
that could arise from the displacement of the grain's charge centroid
from its center of mass. We assume that this dipole moment and the
magnetic dipole moment are oriented randomly, thus giving the grain a
total dipole moment of

\begin{equation}
\label{eq:fe_mag}
\mu^2 = \mu_m^2 + \left(\epsilon a_x Z
  e\right)^2
~~~,
\end{equation}
where we adopt $\epsilon = 0.01$ and $a_x = a$ as for the silicate grains. For a
grain with $Z = 1$ and $a = 4.5$\,\AA, the electric dipole moment due
to the charge distribution (proportional to $Za$) is a factor of four
smaller than its magnetic moment (proportional to $a^3$). Thus, the
electric dipole moment due to the grain charge distribution is a small
correction for all grain sizes of interest.

\subsubsection{Electric Dipole Moment of Impure Fe Nanoparticles}
\label{sec:fe_elec}
While pure quasi-spherical iron nanoparticles seem unlikely to have
electric dipole moments even if charged, there may be nanoparticles
that are predominantly Fe but that also contain non-Fe atoms, such as
C or O.  If these are distributed more or less randomly, it is
plausible that the predominantly Fe nanoparticle could have an
electric dipole moment of several Debye.

To explore this effect, we also compute the rotational emission from a
population of iron grains assuming that a grain of $N_{at}$ atoms has an rms
electric dipole moment of $\beta\sqrt{N_{at}}$\,D, as we assumed for
silicate grains, in addition to its magnetic dipole moment and
electric dipole moment due to its charge distribution. Thus

\begin{equation}
\label{eq:fe_e}
\mu^2= \mu_m^2 + \mu_i^2 + \left(\epsilon a_x Z
  q_e\right)^2
~~~,
\end{equation}
where we average the electric dipole moment $\mu_i$ over the
distribution given by Equation~\ref{eq:dipole_dist} and adopt $\beta =
0.3$\,D.

\subsection{Size Distribution}
\label{subsec:size_dist}
Following previous studies modeling spinning dust emission
\citep[e.g.][]{Draine+Lazarian_1998a, Hoang+Vinh+QuynhLan_2016}, we consider a
log-normal grain size distribution with

\begin{equation}
\label{eq:size_dist}
\frac{1}{n_{\rm H}} \frac{{\rm d}n}{{\rm d}a} =
  \frac{B}{a}{\rm exp}\left\{-\frac{1}{2}\left[\frac{{\rm
            ln}\left(a/a_0\right)}{\sigma}\right]^2\right\}
~~~,
\end{equation}
where the parameters $a_0$ and $\sigma$ determine the peak size and
width of the size distribution. $B$ is a normalization constant
given by

\begin{align}
\label{eq:size_b}
B &= \frac{3}{\left(2\pi\right)^{3/2}}\frac{{\rm
    exp}\left(-4.5\sigma^2\right)}{a_0^3\rho\sigma} \times \nonumber
  \\ 
& \frac{m_{\rm X}b_{\rm X}}{1 + {\rm
  erf}\left[3\sigma/\sqrt{2} + {\rm ln}\left(a_0/a_{\rm min}\right)/\sigma\sqrt{2}\right]}
\end{align}
for $a > a_{\rm min}$ and zero otherwise, where $\rho$ is the grain mass density, $m_{\rm X}$ is the grain mass per
atom of element X, and $b_{\rm X}$ is the number of X atoms per H
consumed by this grain population.

The very smallest grains are subject to rapid sublimation by the
interstellar radiation field. \citet{Guhathakurta+Draine_1989} found that
silicate grains of fewer than 37 atoms ($a \lesssim 4.4\,$\AA) are
photolytically unstable in the local interstellar radiation field
\citep[assuming the spectrum of][]{Mathis+Mezger+Panagia_1983}, and thus we truncate the silicate
grain size distribution at $a_{\rm min} =$ 4.5\,\AA. Likewise,
\citet{Hensley+Draine_2016a} found that iron grains could persist down
to a radius of 4.5\,\AA, and so we adopt $a_{\rm min} =$ 4.5\,\AA\ for
the metallic iron grains as well.

\section{Spinning Dust Emissivity}
\label{sec:spdust}
In this Section, we assess whether the observed AME can be reproduced by spinning
nanosilicate or iron grains.

\subsection{Methodology}
To compute the spinning dust
emission from ultrasmall silicate and iron grains, we employ the \texttt{SpDust} code
\citep{Ali-Haimoud+Hirata+Dickinson_2009,
  Silsbee+AliHaimoud+Hirata_2011}. As the code is
optimized for PAHs, several important changes were necessary. 

First, we have assumed that all silicate and iron grains are spherical
rather than having a population of disklike grains below some radius.

Second, we have employed the formalism of
\citet{Weingartner+Draine_2001b} and \citet{Hensley+Draine_2016a} to
compute the charge distributions of small silicate and iron grains,
respectively. We have made a small change to the charging calculations
of PAHs and silicates by adopting the functional form of $E_{\rm min}$
(the minimum energy at which an electron is capable of tunneling
out of a grain)
suggested by \citet[][Equation~1]{vanHoof+etal_2004}
\citep[cf.][Equation~7]{Weingartner+Draine_2001b}. 

Third, we have calculated the temperature distributions and resulting
infrared emission of both silicate and iron grains following the methods of
\citet{Draine+Li_2001} in order to compute the rotational damping and
excitation due to infrared emission. 

Fourth, \citet{Ali-Haimoud+Hirata+Dickinson_2009} computed the
temperature $T_{\rm ev}$ of atoms evaporating from the surface of PAHs
taking into account the ejection of adsorbed atoms during stochastic heating
events. In this work, we adopt their $T_{\rm ev}$ as a function of
grain size and radiation field intensity for all
grains irrespective of composition. However, unlike
\citet{Ali-Haimoud+Hirata+Dickinson_2009}, we assume that we
are always in the limit that there are available sites on the grain for
atoms to stick (cf. their Equation~60). 

Fifth, the rotational velocity distribution of spinning grains depends
in part on the ability of the grain to rotationally couple to distant
plasma via its dipole moment. However, this coupling is much stronger
to a grain's electric dipole moment than to its magnetic dipole
moment. Thus, when computing the effects of plasma drag, we
consider the magnitude of the grain's electric dipole moment
only. Likewise, we ignore the effect of the grain's magnetic dipole
moment on the trajectory of colliding ions \citep[see][Equations~90
and 91]{Ali-Haimoud+Hirata+Dickinson_2009}.

Finally, we have updated the relevant parts of
the code with the material properties of silicate and iron (e.g. mass
density, number of atoms per grain mass, index of refraction,
etc.).

We investigate the spinning dust emission in a variety of interstellar
environments-- the Cold Neutral Medium (CNM), Warm Neutral Medium
(WNM), Warm Ionized Medium (WIM), reflection nebulae (RN), and
photodissociation regions (PDRs). We adopt the idealized physical
parameters of these environments proposed by
\citet{Draine+Lazarian_1998a}, which we list in Table~\ref{table:ism}.

\begin{deluxetable*}{lccccc}
      \tablecaption{Idealized ISM Phases \label{table:ism}}
    \tablehead{& \colhead{CNM} & \colhead{WNM} & \colhead{WIM} &
      \colhead{RN} & \colhead{PDR}}
    \startdata
    $n_{\rm H}\,({\rm cm}^{-3})$ & 30 & 0.4 & 0.1 & $10^3$ & $10^5$ \\
    $T_g\,({\rm K})$ & 100 & 6000 & 8000 & 100 & 300 \\
    $\chi$ & 1 & 1 & 1 & 1000 & 3000 \\
    $x_{\rm H} \equiv n({\rm H}^+)/n_{\rm H}$ & 0.0012 & 0.1 & 0.99 &
    0.001 & 0.0001 \\
    $x_{\rm C} \equiv n({\rm C}^+)/n_{\rm H}$ & 0.0003 & 0.0003 &
    0.001 & 0.0002 & 0.0002\\
    $y \equiv 2n({\rm H}_2)/n_{\rm H}$ & 0 & 0 & 0 & 0.5 & 0.5\\
    $\gamma$ & 0 & 0 & 0 & 0.1 & 0.1
    \enddata
    \tablecomments{Parameters for the idealized ISM models of
      \citet{Draine+Lazarian_1998b}: hydrogen number density $n_{\rm H}$, gas
      temperature $T_g$, radiation intensity parameter $\chi$, H$^+$
      abundance $x_{\rm H}$, C$^+$ abundance $x_{\rm C}$, H$_2$
      abundance $y$, and H$_2$ formation efficiency $\gamma$. Grains are
    assumed to be illuminated by a radiation field with spectrum given
  by \citet{Mathis+Mezger+Panagia_1983} multiplied by a
  frequency-independent factor $\chi$.}
\end{deluxetable*}

\subsection{Comparison to Other Studies}
Our calculations of the emission from spinning silicate and iron
nanoparticles are in many ways complementary to those of
\citet{Hoang+Vinh+QuynhLan_2016} and \citet{Hoang+Lazarian_2016},
respectively, as they are based on somewhat 
different approaches and assumptions. 

First, by using the \texttt{SpDust} code, we are
employing the Fokker-Planck equation to compute the rotational
velocity distribution of small grains. In contrast, the aforementioned studies
employ an approach based on the Langevin equation that was modified in
order to incorporate the effects of stochastic impulsive torques.
However, \citet{Hoang+Draine+Lazarian_2010} found that impulsive
events had only a minor effect, slightly raising the peak frequency
and slightly extending the high frequency tail of the emission
spectrum. Hence our Fokker-Planck treatment is a good approximation.

Second, \citet{Hoang+Lazarian+Draine_2011} demonstrated
that the spinning dust spectrum of PAHs averaged over a realistic
temperature distribution was nearly identical to one in which the dust
temperature was fixed to 60\,K. Thus, \citet{Hoang+Vinh+QuynhLan_2016}
adopt $T_d = 60$\,K for silicate grains while
\citet{Hoang+Lazarian_2016} compute spinning dust spectra of iron
grains with $T_d = 20$ and 40\,K. In contrast, we solve
for the full temperature distribution and resultant infrared emission
from these grains, which affects the magnitude of the infrared
damping. For a 1\,nm silicate grain, \citet{Hoang+Vinh+QuynhLan_2016}
find IR damping and excitation coefficients of $F_{\rm IR} = 0.6$ and
$G_{\rm IR} = 0.7$, respectively. In contrast, we find $F_{\rm IR} =
3.0$ and $G_{\rm IR} = 0.7$.

Third, those studies adopted $a_{\rm min} = 3.5$\,\AA,
i.e., the same as PAHs, whereas we employ $a_{\rm min} = 4.5$\,\AA\
based upon the sublimation rates for silicate and iron grains
\citep{Guhathakurta+Draine_1989, Hensley+Draine_2016a}.

Finally, unlike those studies, we do not employ observational
constraints based on polarization as ultrasmall grains are expected to
be unaligned \citep{Draine+Hensley_2016a}. Instead, we calculate the
infrared emission and ultraviolet extinction from populations of
nanosilicate and iron grains that could account for the observed AME,
and then compare to observations. Nevertheless,
as we demonstrate below, the conclusions reached regarding the AME
emissivity of iron and silicate nanoparticles are qualitatively
similar.

\subsection{Results}
\begin{figure*}
    \centering
        \scalebox{1.0}{\includegraphics{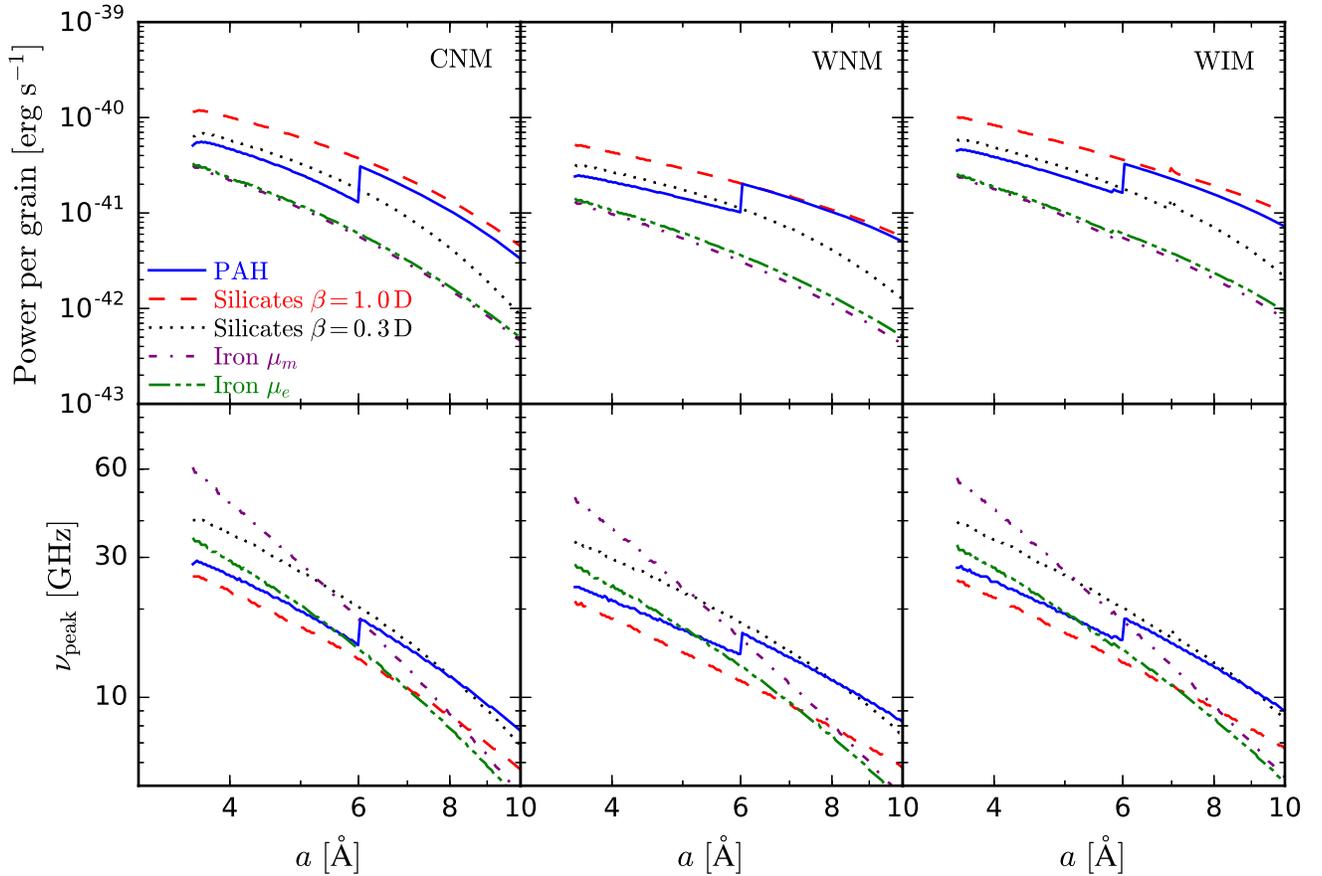}}  
    \caption{{\it Top panels}: A comparison of the rotational power emitted per grain as a
      function of grain size for PAHs, nanosilicates, and iron
      nanoparticles in CNM (left), WNM (middle), and WIM (right)
      conditions. {\it Bottom panels}: A comparison of the peak frequency of the spinning dust
      emissivity $j_\nu$ as a function of grain size for PAHs, nanosilicates,
      and iron nanoparticles in CNM (left), WNM (middle), and WIM
      (right) conditions. We use ``$\mu_m$'' and ``$\mu_e$'' to denote
      pure iron nanoparticles with a total dipole moment
      given by Equation~\ref{eq:fe_mag} and impure iron nanoparticles with a total
      dipole moment given by Equation~\ref{eq:fe_e}, respectively.} \label{fig:emiss_peak} 
\end{figure*}

In the top panels of Figure~\ref{fig:emiss_peak}, we compare the
spinning dust power per grain
as a function of grain size for PAHs, nanosilicates, and iron
nanoparticles in the CNM, WNM, and WIM interstellar environments of
Table~\ref{table:ism}. It is evident that spinning dust emission is strongly
sensitive to the grain size distribution and, in particular, the
abundance of the very smallest grains.

In the bottom panels of Figure~\ref{fig:emiss_peak}, we compare the peak frequency of the
spinning dust emissivity $j_\nu$ of each grain size for each interstellar
environment. Again, strong dependence on grain size is observed. 

\begin{figure*}
    \centering
        \scalebox{0.9}{\includegraphics{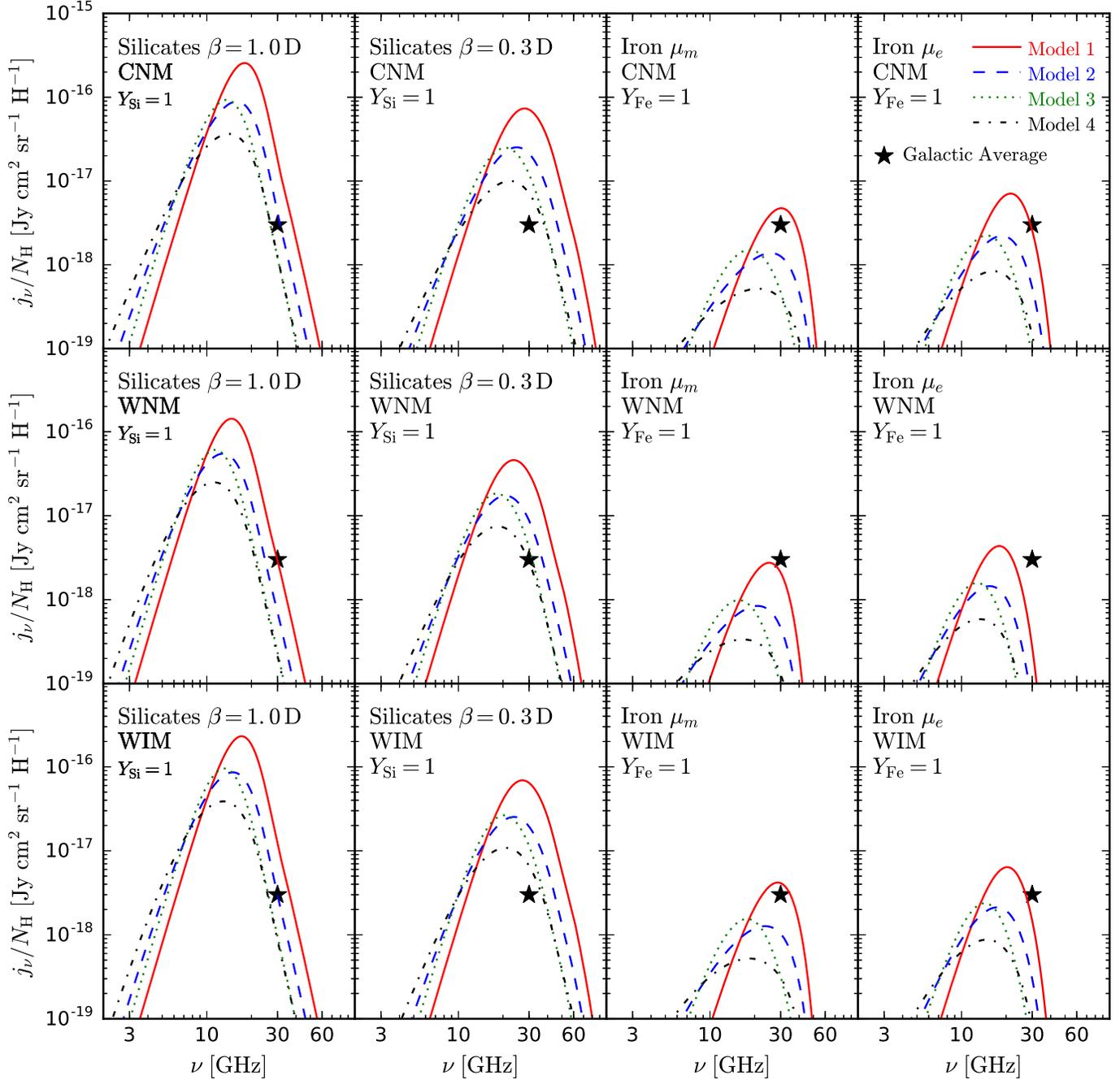}}  
    \caption{The spinning dust SEDs of nanosilicates with $\beta =
      1.0\,{\rm D}$ (first column), nanosilicates with $\beta = 0.3\,{\rm D}$
      (second column), pure iron nanoparticles with a total dipole moment
      given my Equation~\ref{eq:fe_mag}
      (third column), and impure iron nanoparticles with a total
      dipole moment given by Equation~\ref{eq:fe_e} (fourth column)
      for the size distributions defined in
      Table~\ref{table:models} with $Y = 1.0$ and for CNM, WNM, and
      WIM environments. The observed Galactic spinning dust emissivity at 30\,GHz
    is approximately
    $3\times10^{-18}$\,Jy\,sr$^{-1}$\,cm$^2$\,H$^{-1}$ and is indicated by
    the star symbol. Several models of nanosilicate grains (model 1 for
    $\beta = 1.0$\,D and models 1 and 2 for $\beta = 0.3$\,D) can
    account for the entirety of the AME signal with $Y_{\rm Si} <
    20$\%. In contrast, for pure iron, only the model concentrating most of the
    interstellar iron in the smallest nanoparticles (model 1) is able
    to produce a comparable amount of AME, but
    consumes nearly two thirds of the interstellar iron. All models of impure
    iron grains with an appreciable electric dipole moment produce
    insufficient 30\,GHz emission.} \label{fig:spdust} 
\end{figure*}

Given the strong dependencies on grain size, we consider several grain
size distributions for each grain material, which we summarize in
Table~\ref{table:models}. In brief, we adopt $a_0 \in \{a_{\rm min},
  6\,{\rm \AA}\}$ and $\sigma \in \{0.1, 0.3\}$. We plot the resulting
  spinning dust spectrum for each size distribution 
  in Figure~\ref{fig:spdust}, where we have set $Y_{\rm Si}$ and
  $Y_{\rm Fe}$ to 1.0 to obtain a maximum total emissivity for each
  combination of composition and size distribution.

\begin{deluxetable*}{ccccccccc}
      \tablewidth{0pc}
      \tablecaption{Grain Size Distributions\label{table:models}}
    \tablehead{\colhead{Model Number} & \colhead{$a_0$\,[\AA]} &
      \colhead{$\sigma$} & \colhead{$Y_{\rm Si,\,1.0}^{\rm min}$} &
      \colhead{$Y_{\rm Si,\,0.3}^{\rm min}$} & \colhead{$Y_{\rm Si}^{\rm max}$} &
      \colhead{$Y_{{\rm Fe,}\ \mu_m}^{\rm min}$} &
      \colhead{$Y_{{\rm Fe,}\ \mu_e}^{\rm min}$} & \colhead{$Y_{\rm Fe}^{\rm max}$}}
    \startdata
    1 & $a_{\rm min}$ & 0.1 & 0.17 & 0.04 & 0.21 & 0.63 & 1.18 & 0.46 \\
    2 & $a_{\rm min}$ & 0.3 & 0.73 & 0.14 & 0.15 & 2.38 & 5.67 & 0.34 \\
    3 & 6.0 & 0.1 & 2.51 & 0.25 & 0.13 & 6.40 & 39.3 & 0.30 \\
    4 & 6.0 & 0.3 & 2.50 & 0.42 & 0.18 & 7.69 & 21.1 & 0.38
    \enddata
    \tablecomments{Parameters for the grain size distributions of
      models considered in this work. We adopt $a_{\rm
        min} = 4.5\,$\AA\ for both silicate and iron grains. We determine $Y^{\rm min}$ by requiring the
      30\,GHz spinning dust emissivity of that size distribution to be
      $3\times10^{-18}$\,Jy\,sr$^{-1}$\,cm$^2$\,H$^{-1}$ for CNM conditions. $Y_{\rm
        Si,\,1.0}^{\rm min}$ and $Y_{\rm Si,\,0.3}^{\rm min}$ denote
    $Y^{\rm min}$ for nanosilicate grains with $\beta = 1.0$ and
    0.3\,D, respectively. $Y_{{\rm Fe,}\ \mu_m}^{\rm min}$ and $Y_{{\rm Fe,}\ \mu_e}^{\rm min}$
    denote $Y^{\rm min}$ for pure iron nanoparticles with a total dipole moment
      given by Equation~\ref{eq:fe_mag} and impure iron nanoparticles with a total
      dipole moment given by Equation~\ref{eq:fe_e}, respectively.
    We determine $Y^{\rm max}$ by requiring the mid-infrared emission from grains
    with the specified size distribution to reproduce the observed 10\,$\mu$m emission for silicates and the 25\,$\mu$m
    emission for iron grains.}
\end{deluxetable*}

We note that the emissivities obtained for populations of silicate and
iron nanoparticles are similar (within a factor of a few) to those derived by
\citet{Hoang+Vinh+QuynhLan_2016} and \citet{Hoang+Lazarian_2016},
respectively when employing the same size distribution and value of
$Y$. However, when adopting the more
physically-motivated minimum grain size of $a_{\rm min} = 4.5$\,\AA\
(instead of 3.5\,\AA, the same as for PAHs), we derive somewhat lower
emissivities. Indeed, for models of silicate grains with $\beta =
0.3$\,D in CNM conditions and $\sigma = 0.1$, we find a factor of two
increase in 30\,GHz emissivity when adopting $a_0 = a_{\rm min} =
3.5$\,\AA\ versus 4.5\,\AA. Likewise, the small changes in the grain
size distribution represented by models 1 - 4 differ by roughly an
order of magnitude in peak emissivity. This further underscores the
sensitivity of the spinning dust emissivity to the abundance of the
smallest grains.

\begin{figure}
    \centering
        \scalebox{0.9}{\includegraphics{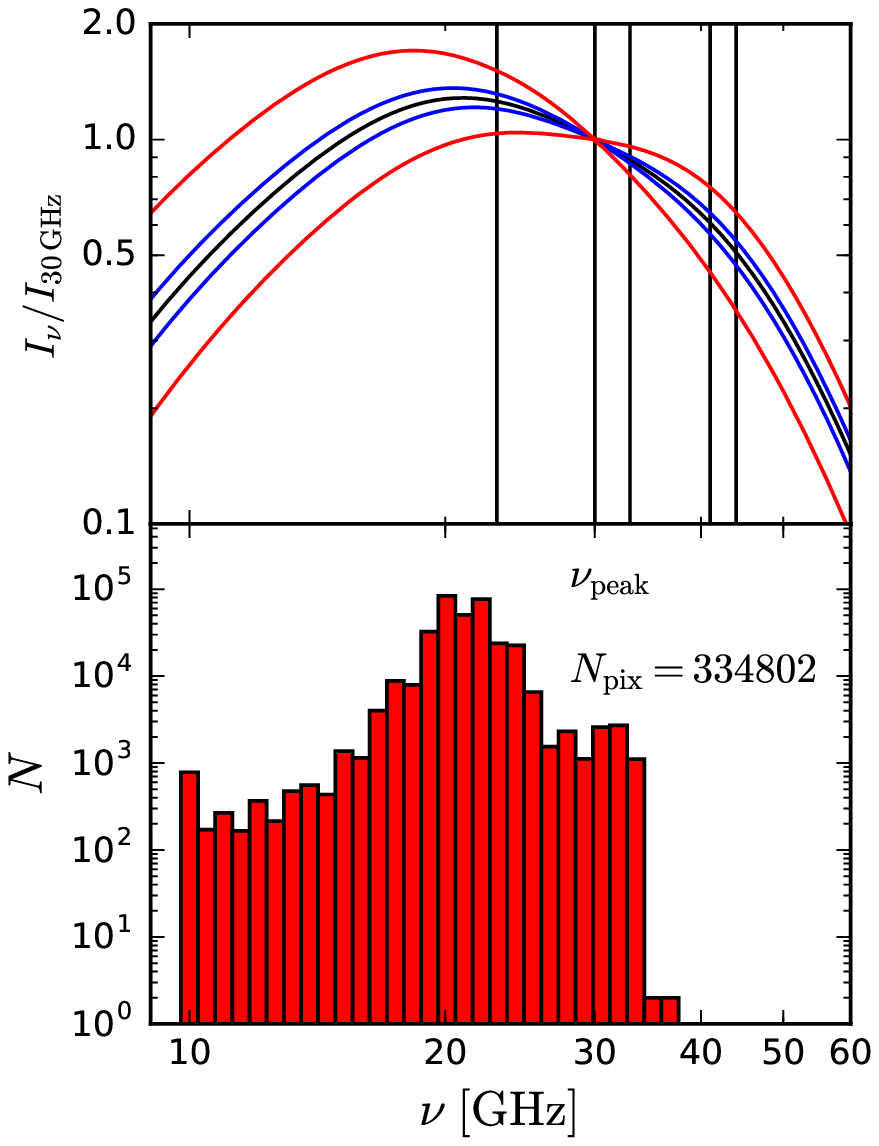}}  
    \caption{Given the full-sky fits to the AME performed by
      \citet{Planck_2015_X}, in the top panel we plot the median (black), 68\%
      confidence interval (blue), and 95\% confidence interval (red)
      of the AME SEDs normalizing to 1 at 30\,GHz. We note that we
      have masked the Galactic plane within $|b| < 5^\circ$ as well as
      pixels in which the fit AME amplitude was less than five times
      the quoted fit uncertainty. This leaves 334,802 pixels at
      $N_{\rm side} = 256$, about 43\% of the sky. The vertical lines
      indicate the 23, 33, and 41\, GHz WMAP bands and the 30 and 44\,GHz
      {\it Planck} bands. In the
      bottom panel, we plot the histogram of AME peak frequencies in
      these pixels.} \label{fig:ame_sed} 
\end{figure}

The Galactic AME has an observed emissivity of roughly
$3\times10^{-18}$\,Jy\,sr$^{-1}$\,cm$^2$\,H$^{-1}$ at 30\,GHz in both
Galactic clouds and the diffuse ISM
\citep{Dobler+Draine+Finkbeiner_2009, Tibbs+etal_2010, 
  Tibbs+etal_2011, Planck_Int_XV, Planck_Int_XVII}, and we therefore
consider models viable only if they can reproduce this
emissivity. However, observations of AME in the diffuse ISM suggest
that the emission may be peaking closer to 20\,GHz
\citep{MivilleDeschenes+etal_2008, Planck_2015_X}. In
Figure~\ref{fig:ame_sed}, we illustrate the range of fit AME SEDs from
the {\it Planck} \texttt{Commander} analysis \citep{Planck_2015_X},
demonstrating that the 20\,GHz specific intensity is on average
$\simeq 20-30\%$ higher than the 30\,GHz specific intensity. Therefore,
while successful models need still to reproduce the observed
$3\times10^{-18}$\,Jy\,sr$^{-1}$\,cm$^2$\,H$^{-1}$ at 30\,GHz, those
which peak at lower frequencies may be favored. For all models, we
define $Y^{\rm min}$ as the minimum value of $Y$ such that the model
has a 30\,GHz emissivity of
$3\times10^{-18}$\,Jy\,sr$^{-1}$\,cm$^2$\,H$^{-1}$.

From Figure~\ref{fig:spdust}, it is clear that nanosilicate grains are
able to produce considerable rotational emission and may account for
the entirety of the AME. The SEDs for models 1 and 2, which concentrate most of the
grain mass into the smallest nanoparticles, and $\beta=0.3$\,D both
peak near 30 GHz, and can approximately reproduce the observed AME
spectrum with $Y_{\rm Si}=0.04$ or 0.14, respectively. Models with
$\beta = 1.0$\,D peak at lower frequencies due to the enhanced damping
from electric dipole emission. While model 1 with $\beta = 1.0$\,D can reproduce the
observed AME emissivity while consuming only a modest amount of the
interstellar silicon ($Y_{\rm Si} = 0.17$), the remaining $\beta =
1.0$\,D models produce inadequate 30\,GHz emission.

In contrast, iron nanoparticles are able to reproduce the observed
30\,GHz AME without
over-consuming the available iron only for pure iron grains with the
size distribution which 
concentrates the grain mass in the very smallest grains, although even
this size distribution consumes roughly two thirds of the available
iron ($Y_{\rm Fe} = 0.63$). Iron grains may, however, be contributing
a portion of the AME signal particularly at $\sim 20$\,GHz.

The discrepancy in emissivity between the nanosilicate and iron grains arises for
several reasons. First, the iron grains are much denser, giving them a
larger moment of inertia than silicate grains of the same
radius. Second, the abundance of solid phase Si and Fe in the ISM are
roughly equal. However, silicate grains are a mix of Si, O, Mg, and Fe
whereas metallic iron grains are pure Fe. Thus, the number
of nanoparticles is much greater in models with $Y_{\rm Si}=1$ than in
models with $Y_{\rm Fe}=1$.

\section{Infrared Emission}
\label{sec:irem}
Both silicate and iron nanoparticles undergo stochastic heating
and emit at mid-infrared wavelengths. In this Section, we place limits
on the abundance of silicate and iron nanoparticles based on the
latest observational constraints on the mid-infrared emission of the
diffuse ISM. Although the wavelengths under consideration are much
larger than the relevant grain sizes, the infrared emission retains
sensitivity to the size distribution due to the size-dependence of the
transient heating of small grains.

Nanosilicate grains have a strong mid-infrared signature
due to emission in the 9.7\,$\mu$m feature. \citet{Li+Draine_2001}
found that at most 15\% of interstellar
silicon could be in dust with $a < 10\,$\AA\ on the basis of a 4.5 -
11.7\,$\mu$m spectrum of the diffuse ISM taken by the Infrared
Telescope in Space \citep{Onaka+etal_1996}.

More recent spectroscopic observations from {\it Spitzer} and AKARI
allow us to place new constraints tailored to grain populations capable of reproducing the
AME. \citet{Ingalls+etal_2011} employed the {\it Spitzer} Infrared
Spectrograph to obtain a 5.2-38\,$\mu$m spectrum of the translucent
cloud DCld\,300.2–-16.9, which we adopt to typify the diffuse ISM. In
order to extend the observed spectrum to shorter wavelengths, we also
employ the combined AKARI and {\it Spitzer} spectrum of the
star-forming SBb galaxy NGC\,5992 compiled by
\citet{Brown+etal_2014}. To remove the starlight component, we have
subtracted a 5000\,K blackbody from this spectrum. Due to the
uncertainty associated with this subtraction, we do not employ the
starlight-subtracted spectrum at wavelengths shorter than 3\,$\mu$m.

\begin{figure}
    \centering
        \scalebox{1.0}{\includegraphics{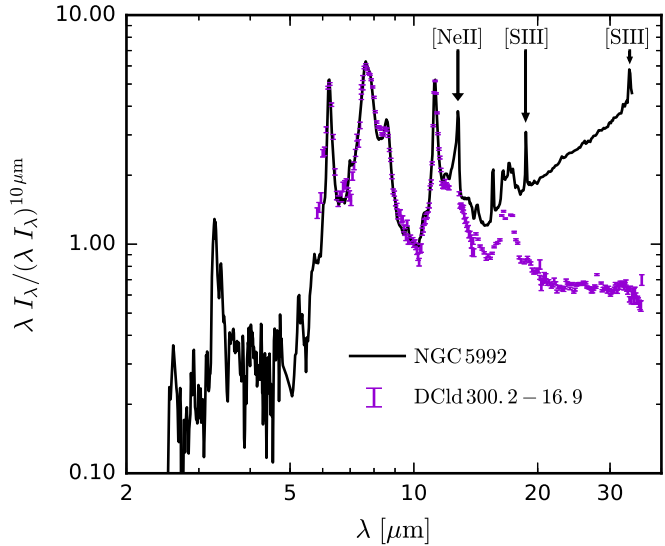}}  
    \caption{A comparison of the mid-infrared spectra of NGC\,5992
      \citep{Brown+etal_2014} after starlight subtraction (black solid
      line) and the Galactic
      translucent cloud DCld 300.2–-16.9
      \citep[][violet data points]{Ingalls+etal_2011}. Both spectra have been normalized to
      1 at 10\,$\mu$m. Given the excellent concordance of the two
      spectral shapes, we employ the combined 3-12\,$\mu$m spectrum as
    a constraint on the mid-infrared dust emission from ultrasmall grains.} \label{fig:pah_compare} 
\end{figure}

The shapes of these two spectra are in
excellent agreement between $\simeq 5 - 12\,\mu$m
(Figure~\ref{fig:pah_compare}), diverging at longer wavelengths presumably because the starlight in
NGC\,5992 is more intense, resulting in higher temperatures for the
big grains.  In addition, the NGC\,5992 spectrum includes line
emission from H{\sc ii} regions, in particular [Ne{\sc ii}]
12.81\,$\mu$m and [S{\sc iii}] 18.71\,$\mu$m, that are of course
absent from the DCld\,300.2–-16.9 spectrum. We scale this spectrum to match the
H{\sc i}-correlated dust emission in the mid-infrared DIRBE bands as
measured by \citet{Dwek+etal_1997}. As the spectrum of NGC\,5992
extends to shorter wavelengths and as the spectrum of DCld\,300.2–-16.9
better matches the shape of the DIRBE SED, we employ the former as a
constraint between 3 and 12\,$\mu$m and the latter longward of
12\,$\mu$m.

We compute the emission from silicate and iron grains by solving for the
temperature distribution function at each grain radius $a$ following
\citet{Draine+Li_2001}. As DIRBE measured dust emission from diffuse,
high-latitude regions, we assume the grains are illuminated by a
radiation field $10^{0.2} \simeq 1.6$ times as intense as the standard
\citet{Mathis+Mezger+Panagia_1983} radiation field. This scaling
factor can be derived from the FIR dust
radiance measured by IRAS and {\it Planck} \citep{Planck_2013_XI} and a standard $R_V = 3.1$
extinction curve \citep[e.g.][]{Fitzpatrick+Massa_2007} using the
fact that power absorbed must equal power radiated and assuming a dust
albedo of $\simeq 0.4$ at optical-UV wavelengths.

\begin{figure*}
    \centering
        \scalebox{0.9}{\includegraphics{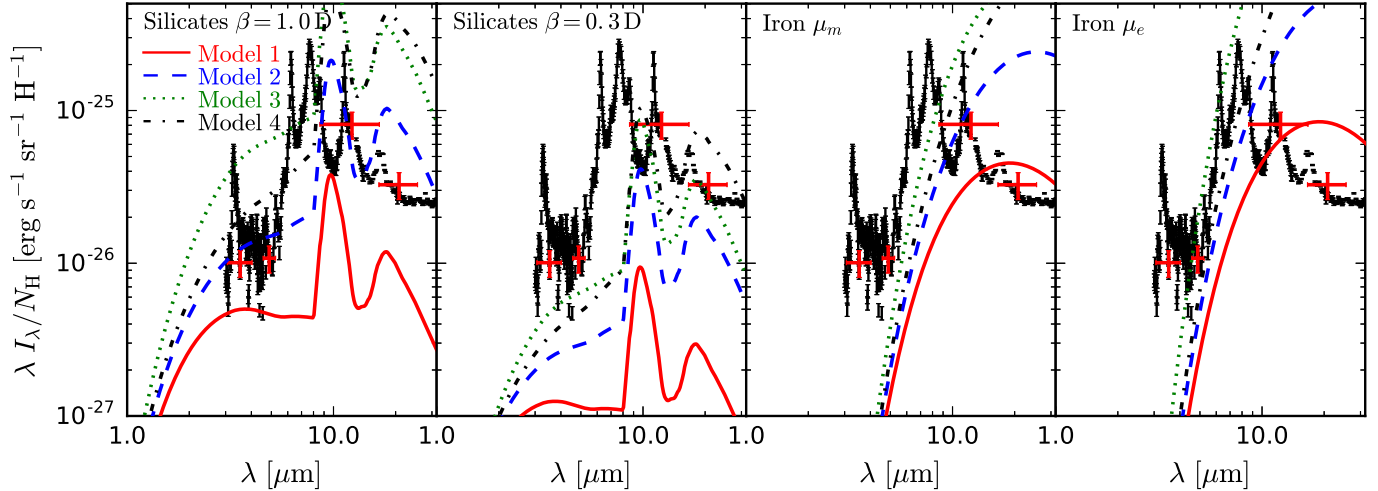}}  
    \caption{In each panel, we plot the combined MIR spectrum of NGC\,5992
      \citep{Brown+etal_2014} and DCld 300.2–-16.9
      \citep{Ingalls+etal_2011} in black, which has been scaled to match
      the H{\sc i}-correlated high-latitude dust emission observed by
      DIRBE \citep[][red error bars]{Dwek+etal_1997}. In the first
      panel, we plot the SEDs of silicate grains with $\beta =
      1.0\,{\rm D}$ and size distributions
    given by Table~\ref{table:models} and with $Y_{\rm Si} = Y_{\rm
      Si,\,1.0}^{\rm min}$. In the second panel, we plot
    the same for silicate grains with $\beta = 0.3\,{\rm D}$ and $Y_{\rm Si} = Y_{\rm
      Si,\,0.3}^{\rm min}$. In the third and fourth panels, we
    plot the same for pure iron
    nanoparticles and impure iron nanoparticles with an appreciable
    electric dipole moment having $Y_{\rm Fe} = Y_{{\rm
      Fe},\ \mu_m}^{\rm min}$ and $Y_{\rm Fe} = Y_{{\rm
      Fe},\ \mu_e}^{\rm min}$, respectively. Silicate grains could produce the entirety of the
    AME without violating observed abundances or mid-infrared
    emission for either value of $\beta$ and the model 1 size
    distribution, whereas all models with a 
    sufficient number of iron nanoparticles to account for the
    observed AME produce more $\simeq20\,\mu$m emission
    than is observed.} \label{fig:irem} 
\end{figure*}

In Figure~\ref{fig:irem}, we compare the observed mid-infrared SED to emission
from nanosilicates and iron nanoparticles where we
have set $Y_{\rm Si}$ and $Y_{\rm Fe}$ to their minimum values (as
given in Table~\ref{table:models}) required to reproduce the entirety
of the Galactic AME at 30\,GHz. We define $Y_{\rm Si}^{\rm max}$ as the value of
$Y_{\rm Si}$ at which the nanosilicates reproduce all of the observed
10\,$\mu$m emission, while we define $Y_{\rm Fe}^{\rm max}$ as the value of
$Y_{\rm Fe}$ at which the iron nanoparticles reproduce all of the observed
emission in the 25\,$\mu$m DIRBE band. These values are listed in
Table~\ref{table:models}.

Silicate grains can simultaneously reproduce the observed 30\,GHz AME
emissivity while respecting
the constraints on infrared emission only for the size distributions
concentrating the grain mass into the smallest nanoparticles (models
1 and 2). In particular, model 1 silicate grains with $\beta = 0.3$\,D and $Y_{\rm
  Si} = Y^{\rm min}$ make only a small contribution to the infrared
emission. In contrast,
none of the models of rotational magnetic dipole emission from iron
nanoparticles can account for all of the AME while respecting
constraints on the observed infrared emission. We thus conclude that if there is a
population of iron nanoparticles in the ISM, their rotational emission
constitutes only a fraction of the observed AME.

\begin{figure}
    \centering
        \scalebox{1.0}{\includegraphics{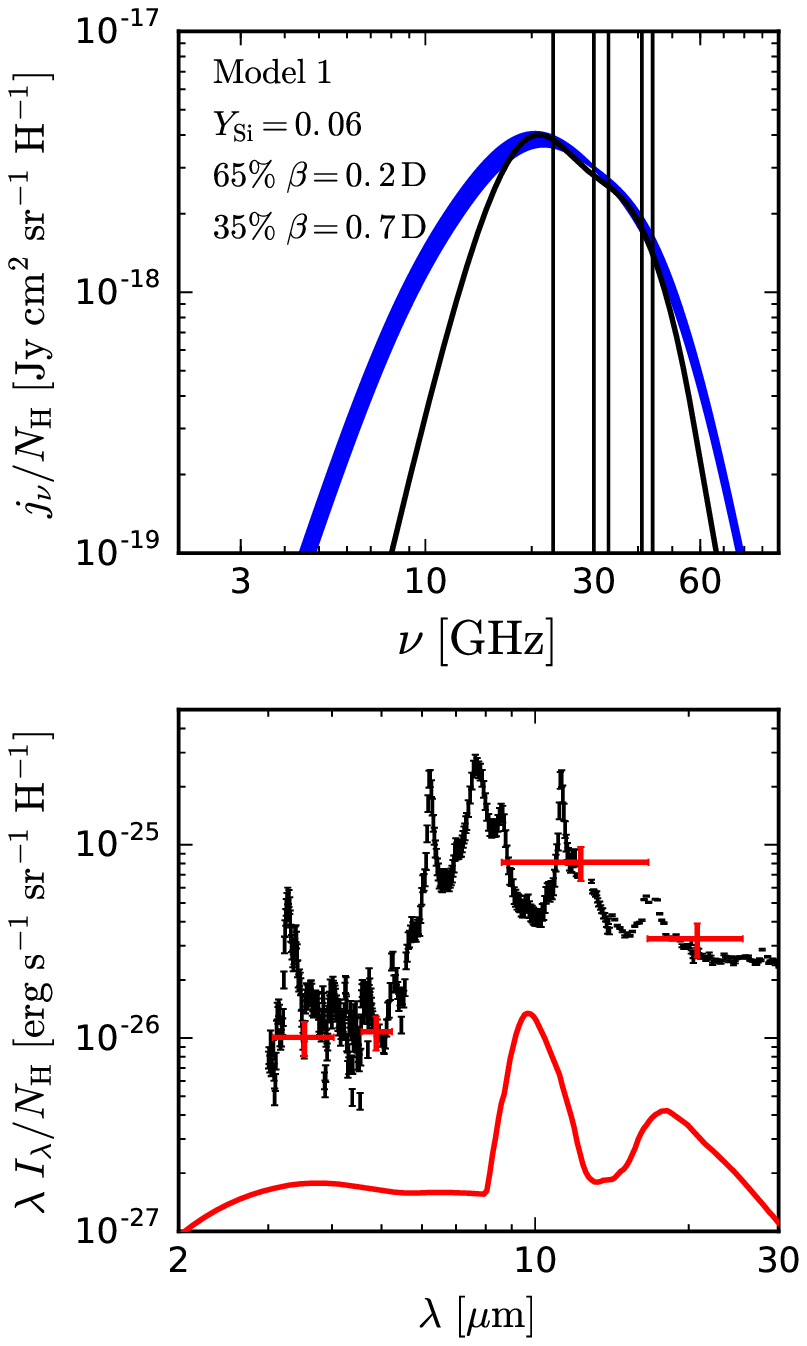}}  
    \caption{In the upper panel, we plot the 68\% confidence interval of the normalized
      \texttt{Commander} AME SEDs which here we normalize to
      $3\times10^{-18}$\,Jy\,sr$^{-1}$\,cm$^2$\,H$^{-1}$ at 30\,GHz
      (blue shaded region). We also plot a model of nanosilicate
      grains having a model 1 size distribution with $Y_{\rm Si} =
      0.06$, 65\% of the grains having an electric dipole moment given
      by Equation~\ref{eq:e_dipole} and $\beta = 0.2$\,D, and 35\% of
      the grains having $\beta = 0.7$\,D. The vertical lines
      indicate the 23, 33, and 41\, GHz WMAP bands and the 30 and 44\,GHz
      {\it Planck} bands. In the lower panel we compare the IR
      emission from this population of nanosilicates (red solid) to
      the observational constraints as in Figure~\ref{fig:irem}.} \label{fig:comm_fit} 
\end{figure}

An example nanosilicate model that would be compatible with the
median AME peak frequency of $\simeq 22$\,GHz and the observed AME emissivity
for CNM conditions has a model 1 size distribution with $Y_{\rm Si} =
0.06$ and electric dipole moments given by
Equation~\ref{eq:e_dipole} with 65\% of the grains having
$\beta=0.2$\,D and the remaining 35\% having $\beta = 0.7$\,D. The upper
panel of Figure~\ref{fig:comm_fit} shows the microwave emission from
this population. For comparison, we also show the 68\% confidence
interval of the AME spectrum inferred by \citet{Planck_2015_X} (see
Figure~\ref{fig:ame_sed}). Note that the AME SED inferred by {\it
  Planck} is really only constrained at 23, 30, 33, 41, and 44\,GHz, the
frequencies where WMAP and {\it Planck} observations were made.

The infrared emission for this example is shown in the lower panel of
Figure~\ref{fig:comm_fit} and falls well below the observational
constraints. We emphasize that the model in Figure~\ref{fig:comm_fit}
is only presented as an example where the nanosilicates account for
all of the AME. In reality, it seems likely that several nanoparticle
components (silicates, PAHs, Fe) will each account for a fraction of
the AME.

\section{Ultraviolet Extinction}
\label{sec:uv_ext}
\begin{figure}
    \centering
        \scalebox{1.0}{\includegraphics{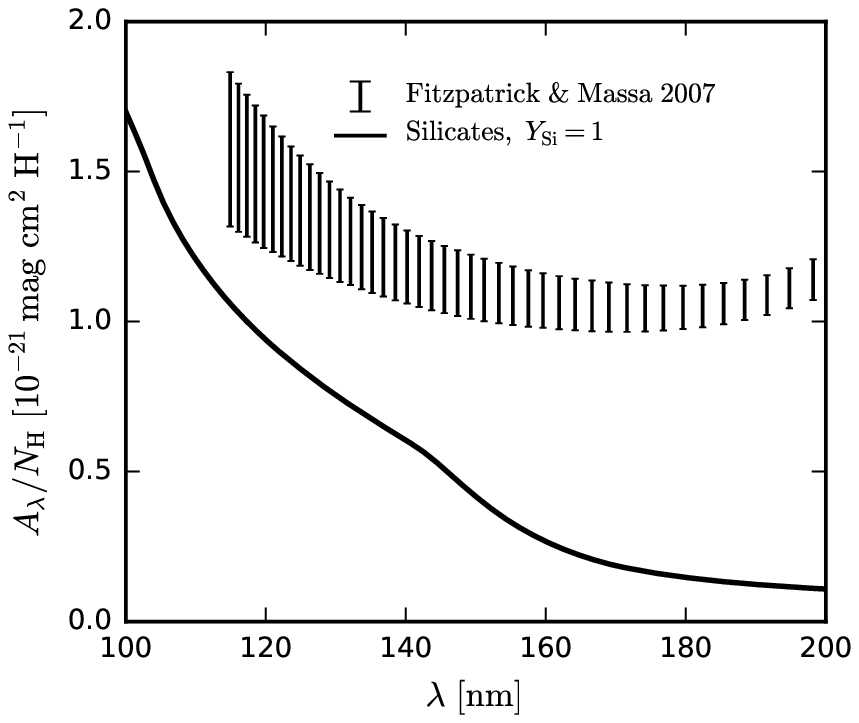}}  
    \caption{We compare the average interstellar extinction curve of
      \citet{Fitzpatrick+Massa_2007} to the extinction arising from a
      population of interstellar nanosilicates with $Y_{\rm Si} =
      1$. To normalize the extinction curve to the hydrogen column, we
    have taken $N_{\rm H}/E(B-V) =
    7.7\times10^{21}$\,cm$^{-2}$\,mag$^{-1}$ \citep{Liszt_2014a,
      Liszt_2014b,Planck_Int_XXIX}. Without assumptions on the
    extinction contributed by larger silicate grains and grains of
    other compositions, the UV extinction curve is unable to constrain
  the fraction of interstellar silicon in the form of nanosilicates.} \label{fig:extcrv} 
\end{figure}

A second constraint on the abundance of interstellar nanoparticles is
the ultraviolet extinction. \citet{Fitzpatrick+Massa_2007} derived a mean Galactic
extinction curve from UV to IR wavelengths based on a sample of 243
stars. To normalize the extinction to the hydrogen column, we take
$N_{\rm H}/E(B-V) = 7.7\times10^{21}$\,cm$^{-2}$\,mag$^{-1}$,
consistent with studies based recent high-latitude H{\sc i} surveys
and reddening maps \citep{Liszt_2014a, Liszt_2014b,Planck_Int_XXIX}. We plot this curve in
Figure~\ref{fig:extcrv} and adopt it as our benchmark. 

\citet{Draine+Hensley_2013} demonstrated that 100\% of the
interstellar iron could be in the form of $a = 5$\,nm Fe nanoparticles
without making a significant contribution to the interstellar extinction
(see their Figure~11). As optical-UV wavelengths are much larger than
the grain sizes of relevance for spinning dust emission
(i.e. the Rayleigh limit), the total extinction will
be insensitive to the nanoparticle size distribution. Thus, we conclude
that the UV extinction provides no additional constraints on the
abundance of iron nanoparticles than were derived in
Section~\ref{sec:irem}.

\citet{Li+Draine_2001} considered the contributions of ultrasmall
silicate grains to the UV extinction curve by adding additional
ultrasmall silicate grains to the dust model of
\citet{Weingartner+Draine_2001}. They found that models which
included an additional $\Delta Y_{\rm Si} \gtrsim 20\%$ of ultrasmall
silicate grains exceeded the observed extinction at $\lambda^{-1}
\gtrsim 7\,\mu$m$^{-1}$. 

In Figure~\ref{fig:extcrv}, we plot the total extinction from a population
of silicate nanoparticles with $Y_{\rm Si} = 1$. At the shortest
wavelengths with observational data, the ultrasmall silicates can only
contribute about two thirds of the total extinction. Without
self-consistently modeling contributions from larger silicate grains
as well as grains of other compositions, it is not possible to derive
more stringent limits on the amount of silicon in nanoparticles than
the $Y^{\rm max}_{\rm Si}$ derived in Section~\ref{sec:irem} on the
basis of infrared emission.

\section{A Generalized Model of Spinning Dust}
\label{sec:general}
We have demonstrated that the AME could arise from spinning
nanoparticles of one or more different compositions, including PAHs,
silicates, and metallic iron. The space of possible spinning dust SEDs
is thus quite large, and it would be useful to articulate generalized,
carrier-independent predictions that can test our theoretical
understanding of spinning dust emission. In this section, we first
highlight three major uncertainties in modeling spinning dust
emission-- the grain size distribution, the distribution of electric
or magnetic
dipole moments, and the grain charge distribution. We then discuss
robust predictions from modeling in light of these
uncertainties, with particular attention to extreme environments like
PDRs. Finally, we suggest observational tests for identifying
spinning dust emission with a particular carrier.

\subsection{Model Uncertainties}
\subsubsection{Grain Size Distribution}
Recognizing that the infrared dust emission features and the AME could
potentially both be explained by a large
population of ultrasmall carbonaceous grains (i.e. PAHs),
\citet{Draine+Lazarian_1998b} proposed a log-normal
component in the size distribution of carbonaceous grains peaked at a
grain radius of $6\,$\AA. Many subsequent studies have since adopted
the log-normal parameterization of the size distribution (as we
discuss in Section~\ref{subsec:size_dist}), which has also been the
default in the \texttt{SpDust} code.

However, absent a detailed theory of dust formation and destruction,
there is no particular reason why the size distribution of ultrasmall
grains should be
log-normal. Given the strong dependence of both the spinning dust
emissivity and peak frequency on grain size (see Figure~\ref{fig:emiss_peak}), we
test the sensitivity of the spinning dust SED to the shape of
grain size distribution. Rather than a log-normal grain size distribution, we
consider truncated power laws of the form

\begin{equation}
\label{eq:size_dist_pow}
\frac{1}{n_{\rm H}} \frac{{\rm d}n}{{\rm d}a} =
  B \left(\frac{a}{a_{\rm min}}\right)^\alpha \quad a_{\rm min} \leq a
  \leq a_{\rm max}
~~~,
\end{equation}
where the normalization constant $B$ is given by

\begin{align}
B &= \frac{3m_{\rm X}b_{\rm X}}{4\pi\rho a_{\rm min}^4} \times
\begin{cases}
\frac{4+\alpha}{-1 + \left(a_{\rm max}/a_{\rm
        min}\right)^{4+\alpha}} & \alpha \neq -4 ,\\
\frac{1}{{\rm ln}\,\left(a_{\rm max}/a_{\rm min}\right)} &
  \alpha = -4.
\end{cases}
\end{align}
$m_X$ and $b_X$ are as defined following Equation~\ref{eq:size_b}.

\begin{figure}
    \centering
        \scalebox{0.9}{\includegraphics{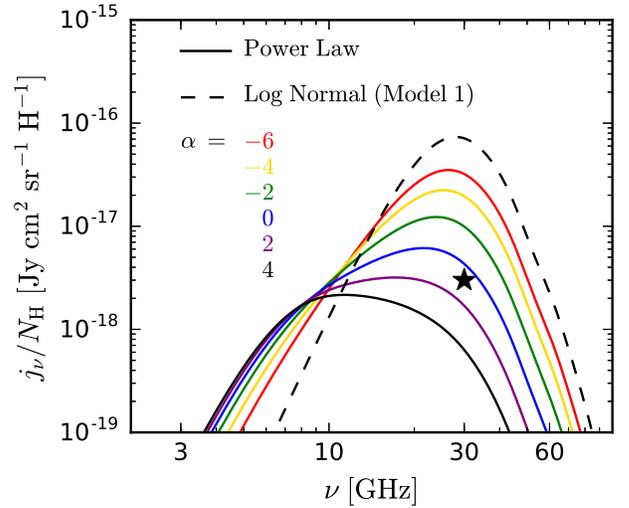}}  
    \caption{Spinning dust emission for silicate grains with $\beta =
    0.3$\,D having a log-normal size distribution (dashed line) and
    power law size distributions (solid lines) assuming $Y_{\rm
      Si} = 1$. For reference, the observed 30\,GHz Galactic spinning dust emissivity of
    approximately $3\times10^{-18}$\,Jy\,sr$^{-1}$\,cm$^2$\,H$^{-1}$
    is indicated with a star. As the power law index
  $\alpha$ increases (i.e. as the size distribution is weighted toward
larger grains) the spinning dust emissivity goes down and the peak
frequency decreases. Further, as $\alpha$ increases, the slope of the
10-30\,GHz portion of the SED becomes shallower. It is evident that
the amplitude, shape, and peak frequency of the spinning dust SED are
all sensitive to the shape of the grain size distribution.} \label{fig:powerlaw_size} 
\end{figure}

In Figure~\ref{fig:powerlaw_size} we plot the spinning dust SED from silicate grains
having electric dipole moment $\beta = 0.3\,$D for different values of
the power law index $\alpha$ ranging from -6 to 4 and assuming $Y_{\rm
  Si} = 1.0$, $a_{\rm min} = 4.5$\AA, $a_{\rm max} = 10\,$\AA, and CNM
conditions. For comparison, we also plot emission from the same
silicate grains having a log-normal size distribution with $a_0 =
a_{\rm min} = 4.5\,$\AA\ and $\sigma = 0.1$ (i.e. model 1, see
Equation~\ref{eq:size_dist}).

As the size distribution becomes weighted toward larger and larger
grains (i.e. as $\alpha$ increases), the total spinning dust emissivity
goes down and the peak frequency shifts to lower frequencies. It is
also notable that the slope of the 10-30\,GHz portion of the spinning
dust SED becomes shallower as $\alpha$ increases, resulting in a less
peaked spectrum. Further, in all cases, the slope of this portion of
the SED differs from that of the log-normal distribution.

We therefore conclude that the spinning dust SED is quite sensitive to
the shape of the grain size distribution, which can influence the
shape, amplitude, and peak frequency of the SED even when all other
parameters are held fixed. As the AME SED becomes better determined
observationally, its ``peakiness'' may provide constraints on the
grain size distribution.

\subsubsection{Dipole Moment Distribution}
The electric dipole moment of a PAH or silicate grain of $N_{at}$ atoms has
been modeled as a ``random walk'' process with each atom contributing
an electric dipole moment $\beta$ in a random direction. This leads to
a Gaussian distribution of electric dipole moments with variance
proportional to $N_{at}$. How sensitive is the resulting spinning dust SED
to the form of this distribution?

\citet{Ali-Haimoud+Hirata+Dickinson_2009} demonstrated that, for a PAH
of radius 3.5\,\AA, the peak frequency of emission could vary from
about 30\,GHz for grains with electric dipole moments twice the rms
value to over 100\,GHz for grains with electric dipole moments less
than 3\% of the rms value (see their Figure 10). This occurs because
the electric dipole radiation provides much of the rotational damping
for $\beta > 0.3$\,D. Further, they found the total radiated power to increase
with increasing electric dipole moment.

\begin{figure}
    \centering
        \scalebox{0.9}{\includegraphics{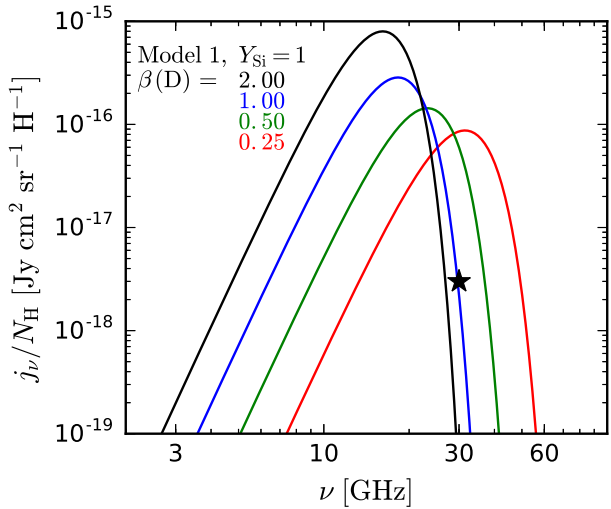}}  
    \caption{Spinning dust emission for silicate grains in which all
      grains are assumed to have a fixed dipole moment per atom
      $\beta$, i.e. there is no 
      averaging over an electric dipole moment distribution. Both the
      spinning dust emissivity and peak frequency are strong functions
    of the electric dipole moment, rendering the spinning dust SED
    sensitive to the shape of the dipole moment distribution. For
    reference, the observed 30\,GHz Galactic spinning dust emissivity
    of approximately
    $3\times10^{-18}$\,Jy\,sr$^{-1}$\,cm$^2$\,H$^{-1}$
    is indicated with a star.} \label{fig:dipole_dist} 
\end{figure}

In Figure~\ref{fig:dipole_dist} we perform a similar analysis for
silicate grains. Assuming a log-normal size distribution with $a_0 = a_{\rm
  min} = 4.5\,$\AA\ and $\sigma = 0.1$ (i.e. model 1), $Y_{\rm Si} =
1$, and CNM conditions, we fix the intrinsic electric dipole moment
for all grains with $N_{at}$
atoms to the value $\mu_i = \beta\sqrt{N_{at}}$ for different values of
$\beta$. The trends observed for PAHs are evident here-- increasing
$\beta$ lowers the peak frequency and increases the emissivity. The
spinning dust SED averaged over electric dipole moments must then
depend on the relative weighting of grains with large dipole moments
that are highly emissive at relatively low frequencies and grains with
small dipole moments that radiate predominantly at higher frequencies.

\subsubsection{Grain Charge Distribution}
A third distribution that must be considered is the grain charge
distribution as many excitation and damping mechanisms, such as
collisions with ions, depend strongly on the grain charge. Detailed
models of grain charging in various interstellar environments have
been presented in the literature \citep{Draine+Sutin_1987,
  Bakes+Tielens_1994, Weingartner+Draine_2001b,
  Hensley+Draine_2016a}, though it should be kept
in mind that these models are often limited by available laboratory
data for the materials of interest and thus come with some
uncertainty. This uncertainty is exacerbated by the difficulty of
constraining the grain charge distribution observationally.

\begin{figure}
    \centering
        \scalebox{0.9}{\includegraphics{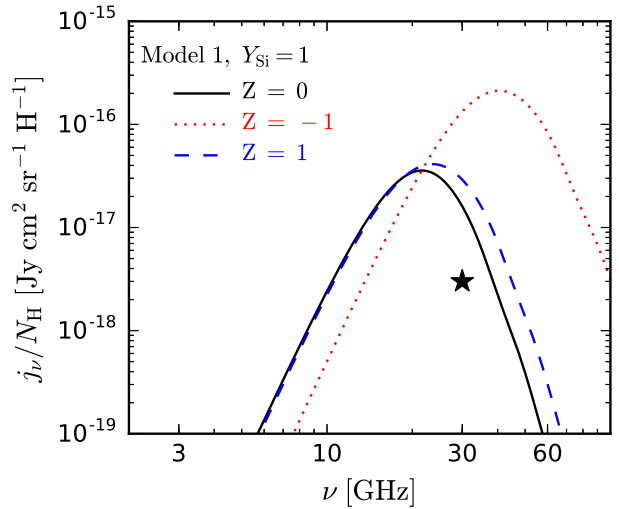}}  
    \caption{Spinning dust emission for silicate grains in which all
      grains are assumed to have a fixed charge $Ze$, i.e. there is no
    averaging over the charge distribution. While neutral and
    positively charged grains behave similarly, negatively charged
    grains are considerably more rotationally excited and have larger
    emissivities and higher peak frequencies. For reference, the
    observed 30\,GHz Galactic spinning dust emissivity of
    approximately $3\times10^{-18}$\,Jy\,sr$^{-1}$\,cm$^2$\,H$^{-1}$
    is indicated with a star.} \label{fig:single_charge} 
\end{figure}

To assess the influence of grain charge on the spinning dust SED, in
Figure~\ref{fig:single_charge} we plot the SEDs of a population of
silicate grains with a log-normal size distribution with $a_0 = a_{\rm min}
= 4.5$\,\AA, $\sigma = 0.1$ (i.e. model 1), $Y_{\rm Si} = 1$, and
$\beta = 0.3$\,D 
assuming CNM conditions and that all grains have charge of either -1,
0, or 1. The SEDs of the positively charged and neutral grains are
quite similar, but the negatively charged grains are significantly
more rotationally excited, having both a higher peak frequency and
emissivity. Particularly in a low-temperature environment like the
CNM, the dominant rotational excitation mechanism is collisions with
gas atoms and ions, with ion collisions being particularly important for
negatively-charged grains. Due to Coulomb attraction,
negatively-charged grains have a much
higher collisional cross section with positive ions than neutral
grains and thus have a rotational excitation coefficient $G_i$
\citep[see][Equation~39]{Draine+Lazarian_1998b} more than a factor of
20 larger.

\citet{Weingartner+Draine_2001b} found that of order 10\% of both PAHs
and ultrasmall silicates are negatively charged in CNM and WNM
conditions, and \citet{Hensley+Draine_2016a} found that roughly
20\% and 50\% of metallic iron nanoparticles are negatively charged in the CNM
and WNM, respectively. These calculations suggest that
negatively-charged grains constitute a non-negligible fraction of
interstellar nanoparticles and their abundance could strongly
influence the shape of the spinning dust spectrum.

\subsubsection{Highly Irradiated Environments}
In the CNM, WNM, and WIM environments we have considered thus far, we
have assumed that the radiation field is identical to the interstellar
radiation field measured by \citet{Mathis+Mezger+Panagia_1983}. In
this section, we consider the more highly irradiated environments of
reflection nebulae (RN) and photodissociation regions (PDRs) where the
radiation field is assumed to be a factor $\chi = 10^3$ and $10^5$
times more intense, resepctively (see Table~\ref{table:ism}).

\citet{Ali-Haimoud+Hirata+Dickinson_2009} demonstrated that the spinning
dust emission spectrum from PAHs in CNM conditions was relatively unaffected by the
strength of the radiation field from $10^{-2} \lesssim \chi \lesssim
10^2$ in either peak emissivity of peak frequency. However, for
$\chi\gtrsim 10^2$, rotational excitation from photon emission
becomes important relative to other excitation mechanisms and both the
peak emissivity and peak frequency begin to increase with increasing
radiation field strength. In the Table~\ref{table:ism} RN environment,
rotational excitation of PAHs is indeed dominated by IR emission,
though collisional excitation is dominant in PDRs
\citep[see][Table~1]{Ali-Haimoud_13}.

\begin{figure*}
    \centering
        \scalebox{0.9}{\includegraphics{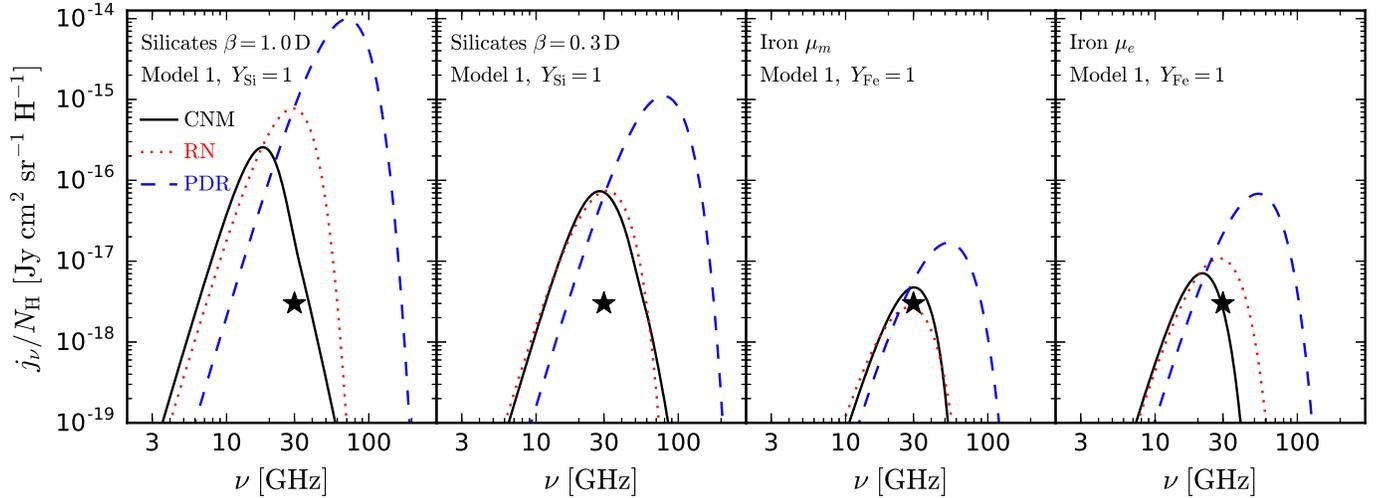}}  
    \caption{Spinning dust emission for silicate and iron grains in
      CNM, reflection nebula (RN), and photodissociation region (PDR)
      environments.} \label{fig:radfield} 
\end{figure*}

Figure~\ref{fig:radfield} illustrates the emission spectra from
silicates and iron grains in CNM, RN, and PDR environments. For
silicates with $\beta = 1.0$\,D, the spectrum shifts to higher peak
frequency and peak emissivity moving from CNM to RN to PDR
environments, much as has been observed for PAHs. However, for
silicates with $\beta = 0.3$\,D, the CNM and RN spectra are remarkably
similar. For silicate grains in both environments, the emission of IR
photons is the dominant rotational {\it damping} mechanism. While 
rotational excitation due to photon emission is the dominant
excitation mechanism for most grain sizes of interest in these
environments, the silicates with the largest electric dipole moments
are also significantly rotationally excited by ion collisions
\citep[see][Equations~90 and
91]{Ali-Haimoud+Hirata+Dickinson_2009}. This additional excitation from
ion collisions accounts for the enhanced 
emissivity and peak frequency of silicate grains with $\beta = 1.0$\,D
in RN conditions relative to those with $\beta = 0.3$\,D or grains in CNM
conditions. In PDRs, collisions with neutrals is by far the dominant
excitation mechanism due to the high density of these regions, and the
peak emissivity and peak frequency of the spinning dust emission
spectrum is accordingly high relative to other environments.

The iron grains tell a similar story-- the enhanced radiation field in
a RN makes little difference to the rotational emission, but the
enhanced gas density of PDRs shifts the spectrum to both higher
emissivity and peak frequency.

\subsection{Generic Predictions of Spinning Dust Theory}
\label{sec:tests}
Spinning dust emission, which may constitute all or part of the
observed AME, could arise from several distinct populations of
nanoparticles, each with their own size, charge, and dipole moment
distributions. Further, particularly in low-resolution CMB
experiments, the emission may be arising from multiple distinct
interstellar environments. All of these factors make predicting the
expected spinning dust spectrum exceedingly difficult. Therefore, fits
to a single AME SED may not be terribly informative as to either the
conditions of the interstellar environment from which the emission
originates or the nature of the dust grains doing the emitting. In
light of this, we suggest a way forward for testing and improving our
theoretical understanding of spinning dust emission by performing
observational tests on the {\it relative} emission spectra of various
environments.

Regardless of the type of nanoparticle producing the rotational
emission, the spinning dust SED is highly sensitive to the abundance
of the very smallest grains. In particularly dense environments where
small grains are expected to be heavily depleted by coagulation, the spinning dust
emissivity per total dust mass (or dust luminosity) should be substantially
smaller than in lower density clouds or the diffuse
ISM. \citet{Tibbs+etal_2016} have presented initial evidence of this
through non-detection of AME in dense Galactic cores. Measurements of the
AME emissivity as a function of the local gas density should reveal a
steady decline in 30\,GHz emissivity with increasing gas density.

A second carrier-independent feature of spinning dust emission is the
importance of ion collisions as a rotational excitation mechanism,
particularly for negatively-charged grains. In environments where
there is an observable change in the ionization fraction of the gas,
such as within a PDR, one may expect a
corresponding shift in the spinning dust emission to higher emissivities
and higher peak frequencies as the gas becomes more
ionized. Note that in photoionized H{\sc ii} gas, the increased
electron density can result in an increased fraction of
negatively-charged nanoparticles with increased rotational excitation
by ion collisions.

In addition to Galactic cores and PDRs, external
galaxies provide an excellent testbed for statistical study of the
dependence of the AME spectra with environmental conditions. The inherent faintness of
the AME signal makes such measurements challenging, but a campaign
targeting nearby galaxies with ample ancillary measurements at other
wavelengths would likely yield invaluable insight into the
environmental factors influencing the AME spectrum and provide tests
of the current formulation of spinning dust theory. 

Finally, as new
datasets such as C-BASS \citep{King+etal_2014} come on-line and component separation
algorithms continue to improve, further analysis of this type will be
made possible on the AME originating in the diffuse Galactic ISM.

\subsection{Identifying the Carrier(s) of Spinning Dust Emission}
Having discussed the generic predictions of spinning dust emission
arising from a nanoparticle of arbitrary composition, we now turn to
ways to identify the particles responsible for the emission. Once
again we argue that it is difficult to extract information from the
spinning dust SED itself, and therefore turn to correlations between
the spinning dust SED and other tracers of ultrasmall grains.

A first test of this type has been tests of the association of the
infrared emission features with the AME, though no compelling
correlation has been found to date
\citep{Tibbs+etal_2011, Vidal+etal_2011, Tibbs+etal_2012,
  Hensley+Murphy+Staguhn_2015, Hensley+Draine+Meisner_2016}. These
studies strongly suggest that the AME does not originate primarily from
the PAH population.

Silicate and metallic iron nanoparticles do not have such striking emission
features, but similar tests may be possible. For instance, if
ultrasmall silicates are able to produce significant 10\,$\mu$m
emission, then there may be observable correlation between this
emission and the AME. However, such measurements will be challenging
due to the difficulty of distinguishing the silicate contribution near
10\,$\mu$m from the wings of the PAH features at 8.6 and 11.2\,$\mu$m.

Very small free-flying Fe nanoparticles become very cold between
photon absorptions and emit relatively little thermal power at FIR and
millimeter wavelengths. However, if Fe nanoparticles are also present
as inclusions, or if the free-flier population includes Fe
nanoparticles with sizes $a > 10$\,\AA, the thermal magnetic dipole
emission can result in spectral flattening of the dust SED at long
wavelengths and a corresponding drop in the polarization fraction of
the thermal dust emission \citep{Draine+Hensley_2013}.  Such
signatures could be correlated with the AME.

 Iron nanoparticles may also
emit strongly near 20\,$\mu$m (see Figure~\ref{fig:irem}). If the Fe
nanoparticles are partially oxidized, energy may be radiated in
vibrational modes (e.g., 22\,$\mu$m FeO, 16.4\,$\mu$m Fe$_3$O$_4$). Assessing
the correlation between the short-wavelength dust emission and the AME
will be valuable if systematic trends do exist, though it will be
difficult to interpret in terms of a specific carrier.

\section{Discussion}
\label{sec:discussion}
Despite the historical association of spinning dust emission with
PAHs, we demonstrate, in qualitative agreement with \citet{Hoang+Lazarian_2016} and
\citet{Hoang+Vinh+QuynhLan_2016}, that spinning nanoparticles of other
compositions can comprise a portion, or even the entirety, of the
observed AME. We demonstrate that variations in the assumed
distributions of grain properties, such as size, charge, and dipole
moments, can lead to large changes in the spinning dust SED. This
suggests that the model space is rich enough to accommodate the
observed diversity in AME SEDs.

However, even the generalized spinning dust hypothesis is not without its
problems. \citet{Hensley+Draine+Meisner_2016} defined a parameter
$f_{\rm PAH}$, based on the ratio of the 12\,$\mu$m intensity to the
total dust radiance, as an indicator of the PAH abundance per dust
mass. They found evidence that this parameter evolved systematically
with both the dust radiance and the dust optical depth, with lower
values of $f_{\rm PAH}$ observed in regions with less dust. This could
be due to an increase in PAH destruction and/or a decrease in PAH
formation in low density regions.  Indeed, \citet{Sandstrom+etal_2010}
found that the PAH abundance in the Small Magellanic Cloud is highest near
molecular clouds, suggesting that PAH formation occurs primarily in the
denser regions and PAH destruction takes place in the more diffuse
regions. Ultrasmall grains of other compositions
are also expected to be more susceptible to destruction in these
conditions \citep{Guhathakurta+Draine_1989}, making $f_{\rm PAH}$ a
seemingly reasonable proxy for the abundance of all ultrasmall grains
irrespective of composition.

Insofar as this is the case, the empirical results of
\citet{Hensley+Draine+Meisner_2016} -- that the AME emissivity per
dust radiance shows no evidence of variation with changes in
environmental conditions -- seem to argue {\it against} the spinning dust 
paradigm in general as spinning dust emission arises preferentially from the very
smallest grains (see Figure~\ref{fig:emiss_peak}). It is possible,
however, that the destruction and formation mechanisms for
PAHs are sufficiently different processes than those of other
grain types that the populations do not vary in tandem. Complementary
tests of the link between the AME and the abundance of ultrasmall
grains of the kind suggested in Section~\ref{sec:tests} are greatly needed.

A second challenge to the spinning dust paradigm is the apparent
positive correlation between the strength of the AME and the strength
of the radiation field \citep{Tibbs+etal_2011, Tibbs+etal_2012,
  Planck_Int_XV, Hensley+Draine+Meisner_2016}. Theoretically, one
expects the radiation field itself
to have only minor effects on the rotational emission in most
environments. Indeed, the radiation field serves mostly to
heat grains to higher temperatures and thereby increase the amount of
rotational damping due to infrared emission
\citep{Draine+Lazarian_1998b, Ali-Haimoud+Hirata+Dickinson_2009,
  Ysard+Verstraete_2010}. 

However, the strength of the radiation field is also correlated with
other environmental parameters, such as the gas density and ionization
state, that are important for the rotational excitation of ultrasmall
grains. Further, regions with a higher radiation field likely have
more star formation, more diffuse molecular clouds, and perhaps more
production of nanoparticles via
fragmentation. Self-consistent ISM models, such as those
constructed by \citet{Ysard+Juvela+Verstraete_2011}, may be helpful in
assessing whether the correlation between the AME and the radiation
field can be reproduced within the spinning dust paradigm.

\section{Conclusion}
\label{sec:conclusions}
The principal conclusions of this work are as follows:

\begin{enumerate}
\item Nanosilicates are viable carriers of the AME if a small fraction
  of the interstellar silicon ($Y_{\rm Si} \sim$0.04 - 0.14 for models 1 and 2 with
  $\beta=0.3$\,D) is in the form of ultrasmall grains. The entirety of
  the AME signal inferred from {\it Planck} and WMAP observations can
  be accounted for by nanosilicates. In Figure~\ref{fig:comm_fit} we present a
  specific model where the AME SED is reproduced with $Y_{\rm Si}=0.06$.
\item Iron nanoparticles are capable of producing some of the
  observed AME, but cannot reproduce it in its entirety without
  violating constraints on the interstellar iron abundance and/or the
  mid-infrared emission.
\item The spinning dust SED is highly sensitive to the
  grain size distribution, charge distribution, and dipole moment
  distribution, making direct inferences from the SED alone
  challenging.
\item Observations constraining
  the variations of the AME spectrum with environment, in particular
  the peak frequency and emissivity per dust column, can provide
  critical tests of the spinning dust hypothesis and help elucidate
  the nature of the AME carrier(s). Variations in the
  AME spectrum as a function of local density and depth into a PDR may
  be especially informative.
\item Some shortcomings of the spinning PAH hypothesis, such as the
  observed link between the AME and the strength of the radiation
  field and the non-correlation of the AME with the PAH abundance,
  also pose problems for spinning non-PAHs. More tests of the spinning
  dust paradigm are needed.
\end{enumerate}

\acknowledgments
{We thank the organizers and participants of the 2016 AME Workshop at
  ESTEC for many stimulating conversations that informed this
  work. BTD acknowledges support from NSF grant
  AST-1408723. The research was carried out in part at the Jet Propulsion
  Laboratory, California Institute of Technology, under a contract
  with the National Aeronautics and Space Administration.}

\bibliography{mybib}

\end{document}